\documentclass[twocolumn,showpacs,aps,floatfix]{revtex4}

\usepackage[latin1]{inputenc}
\usepackage{amsfonts}
\usepackage{amsmath}
\usepackage{amssymb}
\usepackage[american]{babel}
\usepackage{graphicx}
\usepackage{subfigure}
\usepackage{multirow}
\usepackage{psfrag}

\newcommand{\ket}[1]{|#1\rangle}
\newcommand{\bra}[1]{\langle#1|}

\begin{document}

\title{The relativistic quantum channel of communication through field quanta}
\author{M.~Cliche$^{1}$ and A.~Kempf$^{1}$}
\affiliation{$\mbox{}^1$Department of Applied Mathematics,
University of
Waterloo, Waterloo, Ontario, Canada N2L 3G1}

\date{\today}

\begin{abstract}
Setups in which a system Alice emits field quanta which a system Bob receives
are prototypical for wireless communication and have been extensively studied.
In the most basic setup, Alice and Bob are modelled as Unruh-DeWitt detectors
for scalar quanta and the only noise in their communication is due to quantum
fluctuations. For this basic setup we here construct the corresponding
information-theoretic quantum channel. We calculate the classical channel
capacity as a function of the spacetime separation and we confirm that the
classical as well as the quantum channel capacity are strictly zero for
spacelike separations. We show that this channel can be used to entangle Alice
and Bob instantaneously. Alice and Bob are shown to extract this entanglement
from the vacuum through a Casimir-Polder effect.

\end{abstract}
 \maketitle

\section{Introduction}
\label{sec:Channel}

The setup in which two quantum systems, Alice and Bob, communicate using
bosonic field quanta can be viewed as a prototype for wireless communication.
Numerous aspects of this general setup have been studied in the literature, see
e.g. \cite{Caves}. Here, we focus on the most basic case, where Alice and Bob
are modelled as Unruh-DeWitt detectors, i.e., as point-like two-level quantum
systems that interact through a scalar quantum field. Our aim is to construct
and study the information-theoretic quantum channel, $\xi$, i.e., the
completely positive trace preserving map between the input density matrix
$\rho$, in which Alice prepares her detector for the emission, and the output
density matrix $\rho^{\prime}=\xi(\rho)$ of Bob's detector at a later time.
This model captures the communicating of individual q-bits and allows us to
study how communication and entanglement are impacted by both relativity and by
the unavoidable noise that is due to the quantum fluctuations of the field.

Concretely, we construct the quantum channel and provide a perturbative
expansion for it in terms of Feynman-like diagrams. We also calculate the
classical channel capacity of the quantum channel as a function of the
detectors' spacetime separation. We then show, to all orders in perturbation
theory, that both the classical and the quantum channel capacities are strictly
zero when Bob and Alice are spacelike separated. The impossibility of
superluminal signalling has of course been discussed before, see e.g.
\cite{Ahar}. What is new here is that we prove the impossibility of
superluminal signalling information-theoretically by constructing and studying
the quantum channel. We will then discuss how Alice and Bob can use the quantum
channel to extract entanglement from the vacuum. It has been known that Alice
and Bob when coupled to a quantum field can have non-trivial entanglement
dynamics, see e.g. \cite{Hu}. It is also known that, due to the entanglement of
the vacuum \cite{Rezn1,Rezn2}, or the exchange of virtual photons
\cite{Fran08}, two detectors can become entangled even at spacelike
separations, and the speed with which this can happen has been discussed. Here,
we will show that Alice and Bob can naturally and instantaneously become
entangled through the Casimir-Polder effect.

To begin, let us denote the overall Hilbert space by ${\cal H}={\cal
H}^{(1)}\otimes{\cal H}^{(2)}\otimes{\cal H}^{(3)}$, where the first two
Hilbert spaces belong to the detectors of Alice and Bob respectively and where
the third Hilbert space is that of the field. Wherever necessary to avoid
ambiguity we will denote operators $O$ or states $\ket{\psi}$ which live in the
Hilbert space ${\cal H}^{(j)}$ by a superscript (j), for example, $O^{(j)}$ and
$\ket{\psi^{(j)}}$ with $j\in\{1,2,3\}$.  Also, when such operators occur
tensored with identity operators, such as $I^{(1)} \otimes I^{(2)}\otimes
O^{(3)}$, we will often abbreviate this as, for example, $O^{(3)}$. The
Hamiltonian of the system is
\begin{eqnarray}
 &&H=H_{F}+H_{D}+H_{int} \nonumber\\
 &&H_{F}=\int d^3x \left(\frac{1}{2}\pi^{2}(x)+\frac{1}{2}(\nabla\phi(x))^2+\frac{1}{2}m^2\phi^{2} (x)\right) \nonumber \\
 &&H_{D}=\sum_{j=1}^{2}E_e \ket{e^{(j)}}\bra{e^{(j)}} + E_g \ket{g^{(j)}}\bra{g^{(j)}} \nonumber \\
&& H_{int}=\sum_{j=1}^{2}\alpha_j\eta
\Big(\ket{e^{(j)}}\bra{g^{(j)}}+\ket{g^{(j)}} \bra{e^{(j)}}\Big)\phi(x_j)
\label{ty}
\end{eqnarray}
where $H_{F}$ is the Hamiltonian of a free field, $H_{D}$ is the Hamiltonian of
the two detectors, $H_{int}$ is the interaction Hamiltonian between the field
and the detectors, $\alpha_j$ is the coupling constant of the $j$'th detector
($j\in\{1,2\}$), $\phi(x_j)$ is the field at the point of the $j$th detector,
and
$m^{(j)}:=\left(\ket{e^{(j)}}\bra{g^{(j)}}+\ket{g^{(j)}}\bra{e^{(j)}}\right)$
is the monopole matrix of the $j$th detector. The function $\eta(t)$ will be
used to describe the continuous switching on and off of the detectors within
some finite time interval. The use of suitably smooth switching functions
allows one to avoid certain divergences associated with hard on and off
switches, \cite{Satz07}. For simplicity we will always choose the same
switching function $\eta(t)$ for both detectors. We note that the type of
interaction term between the detector and the field that we use in Eq.\ref{ty}
has been extensively studied in the field of quantum field theory in curved
space \cite{Bir}.

The paper is organized as follows. In Sec. II we show that causality is
manifest in the channel.  In Sec. III we study the properties of the channel
and derive a Kraus representation, and in Sec. IV we compute explicitly the
classical channel capacity of the channel. In Sec. V we present a perturbative
expansion of the channel.  In Sec. VI we show that the channel can extract
entanglement from the vacuum and in Sec. VII we compare our channel with
similar models which were analyzed in the quantum optics framework. In the last
section we propose extensions. We work with the natural units $\hbar=c=1$.

\section{Causality}
\label{sec:causality}

The so-called Fermi problem arises in any system that is analogous to two atoms
communicating via the electromagnetic field, and it has been studied
extensively, see e.g. \cite{Pow97}. Consider, in the vacuum, the probability,
$P_{Fermi}$, that a photon is emitted by atom 1 followed by the absorption of a
photon by atom 2. In our model, it is the probability if starting with the
state $\ket{e^{(1)}}\ket{g^{(2)}}\ket{0}$ to end in the state
$\ket{g^{(1)}}\ket{e^{(2)}}\ket{0}$.  Using the perturbative expansion of the
evolution operator $U(t_f,t_i)$ in the interaction picture, one obtains the
transition probability
\begin{eqnarray}
P_{Fermi}&=&\left|\bra{e^{(1)}}\bra{g^{(2)}}\bra{0}U(t_f,t_i)\ket{g^{(1)}}\ket{e^{(2)}}\ket{0}\right|^2 \nonumber \\
&=&\Big|\alpha_1\alpha_2\int^{t_f}_{t_i} dt_1\int^{t_f}_{t_i} dt_2 \eta(t_1)\eta(t_2)\nonumber \\
&& \times e^{i\Delta E(t_2-t_1)}D_{F}(x_1(t_1),x_2(t_2))\Big|^2+ O(\alpha^6)
\nonumber\\ \label{Pfermi}
\end{eqnarray}
where $D_F(x-y):=\bra{0}T\phi(x)\phi(y)\ket{0}$ is the Feynman propagator and
where we defined $\Delta E := E_e-E_g$.  By choosing the separation between the
two detectors $|\vec{x}_{1}-\vec{x}_{2}|$ and a time interval $t_f-t_i$ in which
both detectors are on we can choose the spacetime windows for emission and
absorption to be time-like or spacelike (or mixed) relative to another. The
Fermi problem is the fact that this probability amplitude, from
Eq.(\ref{Pfermi}), is non-vanishing even in the case of spacelike separation.
Technically, this is due to the non-vanishing tail of the Feynman propagator
outside the lightcone. Hegerfeld and Feynman showed that in fact no Feynman
propagator can identically vanish outside the lightcone, \cite{Heger,Dirac}.

This reinforces the need to clarify the reason for the non-vanishing of the
Fermi probability in the spacelike separated case. As was pointed out in
\cite{Fran08}, the key to resolving the puzzle is to take into account that
measurements on the detectors are local measurements. Namely, Bob performs a
measurement only of his detector 2, he does not measure Alice's detector, nor
does he measure the field. This means that Fermi's probability amplitude is the
amplitude for just one of several processes that Bob cannot distinguish. What
should actually vanish for spacelike separations is the sum of the probability
amplitudes for all processes that depend on the state of Alice.

Here, our first aim is to make this argument explicit within the
information-theoretic framework of quantum channels. To this end, we notice
that Bob's ignorance of Alice' and the field's state at the late time $t_f$
means that at $t_f$ both the state of Alice's detector and the state of the
field are to be traced over. These traces perform the sum over the probability
amplitudes for processes that Bob cannot distinguish. We therefore naturally
arrive at the description of a quantum channel $\xi: ~\rho^{(1)}\rightarrow
\xi(\rho^{(1)})=\rho^{(2)\prime}$. Here, the input is the initial density
matrix $\rho^{(1)}$ of Alice at $t_i$ and the output of the channel is Bob's
density matrix $\rho^{(2)\prime}$ at $t_f$.

We assume that the system starts in the state
$\rho(t_i)=\rho^{(1)}\rho^{(2)}\rho^{(3)}$, where the initial state of Alice'
detector, $\rho^{(1)}$, is arbitrary, the initial state of the Bob's detector,
$\rho^{(2)}$, is the ground state and the initial state of the field,
$\rho^{(3)}$, is the vacuum. The full density matrix evolves according to
$\rho(t_f)=U(t_f,t_i)\rho(t_i) U^{\dag}(t_f,t_i)$, where in the interaction
picture $U(t_f,t_i)=T e^{-i\int_{t_i}^{t_f}dt'H_{int}(t')}$. As always, the
time evolution can be formulated in terms of an infinite series of commutators
\cite{Fran02}:
\begin{eqnarray}
\rho(t_f)&=&\rho^{(1)}\rho^{(2)}\rho^{(3)}+\sum_{j=1}^{\infty}\Big((i)^j\int_{t_i}^{t_f}dt_1...\int_{t_i}^{t_{j-1}}dt_j \nonumber\\
&&\times [[...[\rho^{(1)}\rho^{(2)}\rho^{(3)},H_{int}(t_n)],...],H_{int}(t_1)]
\Big). \nonumber \\
\end{eqnarray}
Then, the trace over detector 1 and the field, which we will denote
$Tr_{(1,3)}$, gives the final state $\rho^{(2)}(t_f)=\xi(\rho^{(1)})$ of Bob's
detector:
\begin{eqnarray}
\xi(\rho^{(1)})&=&\rho^{(2)}+\sum_{j=1}^{\infty}\Big((i)^j\int_{t_i}^{t_f}dt_1...\int_{t_i}^{t_{j-1}}dt_j \nonumber\\
&&\times
Tr_{(1,3)}[[...[\rho^{(1)}\rho^{(2)}\rho^{(3)},H_{int}(t_n)],...],H_{int}(t_1)]\Big).\nonumber\\
\label{xi}
\end{eqnarray}
To prove causality from this starting point, we will use the following simple
lemmas:
\begin{description}
            \item[I)] Traces are cyclic and $ Tr\left([A,B]\right)=0$.
            \item[II)] $[A^{(1)} B^{(2)} C^{(3)},D^{(1)} E^{(2)} I^{(3)}] = \\ \Big\{[A^{(1)},D^{(1)}]  \left(B^{(2)}E^{(2)} \right) \\+ \left(D^{(1)}A^{(1)}\right)  [B^{(2)},E^{(2)}]\Big\}  C^{(3)}$.
            \item[III)] $\exists$ $\{R^{(1)}_k,S^{(2)}_k,T^{(3)}_k\}$ such that $[[...[A^{(1)}B^{(2)}C^{(3)},D^{(1)}E^{(2)}],...],F^{(1)}G^{(3)}]= \sum_{k}R^{(1)}_kS^{(2)}_kT^{(3)}_k$.
\end{description}
Now in Eq.(\ref{xi}), the terms that have a dependence on the input
$\rho^{(1)}$ must have at least one $m^{(1)}\phi(x_1)$ which multiplies
$\rho^{(1)}$ since otherwise we simply have $Tr(\rho^{(1)})=1$.  In addition,
since the trace of commutators vanishes (I), the non-vanishing terms which
have a dependence on $\rho^{(1)}$ need to be interacting with at least one
$m^{(2)}\phi(x_2)$, such that all the terms dependent on $\rho^{(1)}$ will be
of the form
\begin{eqnarray}
f_n\left(\rho^{(1)}\right)& =& Tr_{(1,3)}\Big( [[...[\rho^{(1)}\rho^{(2)}\rho^{(3)},m^{(j)}\phi(x_j)],\nonumber \\
& & ...],m^{(r)}\phi(x_r)]\Big) \label{fn}
\end{eqnarray}
where at least one of the indices $\{j...r\}$ is equal to 1 and at least one of
the indices is equal to 2, and $n$ is the number of commutators ($n\geq 2$).
Note that the time dependence is implicit in this formulation, each $\phi(x)$
is integrated over time such that the time  difference between two $\phi(x)$ is
at most $t_f-t_i$.  If the last index in Eq.(\ref{fn}) is 1, using (III) for
everything before the last commutator, and (II) to expand the last commutator,
$f_n\left(\rho^{(1)}\right)$ would simplify to:
\begin{eqnarray}
f_n\left(\rho^{(1)}\right)&=&\sum_k Tr_{(1,3)}\left([R^{(1)}_k S^{(2)}_k T^{(3)}_k,m^{(1)}\phi(x_1)  ]\right) \nonumber\\
&=&\sum_k S^{(2)}_k\Big\{\nonumber\\
&&Tr_{(3)}\left(T^{(3)}_k\phi(x_1)\right)Tr_{(1)}\left([R^{(1)}_k,m^{(1)}]\right)\nonumber\\
&&+Tr_{(3)}\left([T^{(3)}_k,\phi(x_1)]\right)Tr_{(1)}\left(m^{(1)}R^{(1)}_k\right)  \Big\}\nonumber \\
&=&0. \nonumber
\end{eqnarray}
Thus the non-vanishing contributions of $f_n\left(\rho^{(1)}\right)$ must come from commutators for which the very last index is 2.  Now, let us consider the rightmost occurrence of index 1 and let us apply (III) to the commutators to the left of it:
\begin{eqnarray}
f_n\left(\rho^{(1)}\right)&=&\sum_k Tr_{(1,3)}\Big(\nonumber \\
&&[[...[[R^{(1)}_k S^{(2)}_k T^{(3)}_k, m^{(1)}\phi(x_1)],m^{(2)}\phi(x_2) ]\nonumber\\
&&...],m^{(2)}\phi(x_2)] \Big). \label{fn2}
\end{eqnarray}
We can expand the most inner commutators with (II) to obtain:
\begin{eqnarray}
[R^{(1)}_kS^{(2)}_kT^{(3)}_k&,&m^{(1)}\phi(x_1)]=\nonumber\\
&&[R^{(1)}_k,m^{(1)}]\left(S^{(2)}_kT^{(3)}_k \phi(x_1)\right)   \nonumber \\
&&+m^{(1)}R^{(1)}_k\left(S^{(2)}_k[T^{(3)}_k,\phi(x_1)]\right).
\end{eqnarray}
Notice that when the first term is back in Eq.(\ref{fn2}) it forms an expression of the form
\begin{eqnarray}
&&\sum_k Tr_{(1,3)}\Big([R^{(1)}_k,m^{(1)}]\nonumber \\
&&\times[[...[S^{(2)}_kT^{(3)}_k \phi(x_1),m^{(2)}\phi(x_2) ]...],m^{(2)}\phi(x_2)] \Big) \nonumber
\end{eqnarray}
which implies that after the tracing out of detector 1 this term is always absent.  Notice also that when the second term is back in Eq.(\ref{fn2}), it gives an expression of the form:
\begin{eqnarray}
f_n\left(\rho^{(1)}\right)&=&\sum_k Tr_{(1,3)}\Big(m^{(1)}R^{(1)}_k\nonumber \\
&&\times[[...[S^{(2)}_k[T^{(3)}_k,\phi(x_1)],m^{(2)}\phi(x_2)]\nonumber\\
&&...],m^{(2)}\phi(x_2)] \Big).
\end{eqnarray}
Therefore, the term $[T^{(3)}_k,\phi(x_1)]$ will be multiplied on each side by some powers of $\phi(x_2)$, so there exists a set of operators  $V_{k,i,j}^{(2)}$ such that:
\begin{eqnarray}
f_n\left(\rho^{(1)}\right)&=&\sum_{k,i,j}\Big\{V_{k,i,j}^{(2)}Tr_{(1)}\left(m^{(1)}R^{(1)}_k\right)\nonumber\\
&&\times Tr_{(3)}\left(\phi^{i}(x_2)[T^{(3)}_k,\phi(x_1)]\phi^{j}(x_2)\right) \Big\}.\nonumber\\
\end{eqnarray}
Using cyclicity of the trace (I), this expression can be simplified to:
\begin{eqnarray}
f_n\left(\rho^{(1)}\right)&=&\sum_{k,i,j}\Big\{V_{k,i,j}^{(2)}Tr_{(1)}\left(m^{(1)}R^{(1)}_k\right)\nonumber\\
&&\times Tr_{(3)}\left(T^{(3)}_k[\phi(x_1),\phi^{i+j}(x_2)]\right) \Big\}.\nonumber\\
\end{eqnarray}
Note that all the information about $\rho^{(1)}$ is contained in the operators
$R^{(1)}_k$. Causality in the channel therefore follows directly from
microcausality in quantum field theory \cite{Pesk}, namely from the fact that
$[\phi(x),\phi(y)]\vert_{(x-y)^2>0}=0$ (where
$(x-y)^2=-(x^{0}-y^{0})^2+(\vec{x}-\vec{y})^2$). If the two detectors are
spacelike separated during the entire interaction, $\rho^{(2)}(t_f)$ does not
depend on the state $\rho^{(1)}$, i.e., Bob's detector 2 is not sensitive to
the state in which Alice prepared detector 1.

\section{Noise Structure of the channel}
\label{sec:Kraus}

Let us now calculate the precise quantum channel for both time-like and
spacelike separations. Since the evolution of the full system is unitary, our
channel is necessarily described by a CPTP map \cite{Niel00}. Then, as we will
show, assuming detector 2 starts in the ground state,
$\rho^{(2)}=\ket{g^{(2)}}\bra{g^{(2)}}$, we can write the channel map in the
following way, in the basis $\ket{e^{(2)}},\ket{g^{(2)}}$,
\begin{eqnarray}
\xi\left(\left(\begin{smallmatrix} \theta& \gamma\\ \gamma^{\ast}&\beta \end{smallmatrix} \right)\right)&=&\left(\begin{smallmatrix} 0& 0\\ 0&1\end{smallmatrix} \right)+\left(\begin{smallmatrix} P_e& 0\\ 0&-P_e\end{smallmatrix} \right)+\theta \left(\begin{smallmatrix} A& 0\\ 0&-A\end{smallmatrix} \right)   \nonumber \\
&&+\beta \left(\begin{smallmatrix}B& 0\\ 0&-B\end{smallmatrix} \right)
 + \gamma \left(\begin{smallmatrix} 0& C\\ D&0\end{smallmatrix} \right) + \gamma^{\ast}\left(\begin{smallmatrix} 0& D^{\ast}\\ C^{\ast}&0\end{smallmatrix} \right)\nonumber \\ \label{channel}
\end{eqnarray}
where we use $\theta+\beta=1$. All terms are space-time scalars.  Note that
$A,B,C$ and $D$ are causal terms in the sense that they depend on the input
density matrix $\rho^{(1)}$. In contrast, $P_e$ represents noise in the quantum
channel since its presence does not depend on the input $\rho^{(1)}$.  To prove
Eq.(\ref{channel}), we will use the following properties which are easy to
verify ($k\in \mathbb{Z}$):
\begin{description}
            \item[i)] $ Tr\left(\rho^{(3)}\phi^{2k+1}\right)=0$.
            \item[ii)] $m^{2k+1}$ has no diagonal elements, \\
                    and therefore $Tr\left(m^{2k+1} M_d\right)=0$ where $M_d$ is any diagonal matrix.
            \item[iii)] $m^{2k}$ has only diagonal elements, \\
                    and therefore $Tr\left(m^{2k}M_{nd}\right)=0$ where $M_{nd}$ is any matrix with no diagonal elements.
\end{description}
In a series expansion of the non-causal terms, each order has the form
$\rho^{(2)}m^{(2)k}Tr\left(\rho^{(3)}\phi(x_2)^{k}\right)$.  Thus, because of
(i) the non-vanishing terms will be proportional to $\rho^{(2)}m^{(2)2k}$, and
because of (iii)  we know that these are diagonal.  Therefore, because we have
trace preservation and because detector 2 starts initially in the ground state,
there cannot be a more general expression for the non-causal terms of
Eq.(\ref{channel}).  For the causal terms, each order in a series expansion
have the form $\rho^{(2)}m^{(2)k}
Tr(m^{(1)j}\rho^{(1)})Tr\left(\rho^{(3)}\phi(x_1)^{j}\phi(x_2)^{k} \right)$.
Now consider the case where the input density matrix $\rho^{(1)}$ is diagonal,
then because of (ii) the non-vanishing terms will have $j$ even.  Using (i),
this also means we need $k$ to be even, hence $\rho^{(2)}m^{(2)k}$ is diagonal
following (iii).  A similar argument can show that an input density matrix with
no diagonal elements cannot have diagonal elements at the output.  Finally,
trace preservation, hermiticity and linearity of the channel are sufficient
properties to prove the validity of Eq.(\ref{channel}).

From this analysis, we can find a Kraus representation by imposing
$\xi\left(\left(\begin{smallmatrix} \alpha& \gamma \\ \gamma^{\ast}&\beta
\end{smallmatrix} \right)\right)= \sum_{k=1}^{4}E_{k}\left(\begin{smallmatrix}
\alpha& \gamma\\ \gamma^{\ast}&\beta \end{smallmatrix} \right)E_{k}^{\dag}$ and
$\sum_{k=1}^{4}E_{k}^{\dag}E_{k} = I$ where we use
$E_k=\left(\begin{smallmatrix} a_{1k}& a_{2k}\\ a_{3k}&a_{4k} \end{smallmatrix}
\right)$.  Solving this nonlinear system of equations is relatively
straightforward as we have more unknowns than equations, so for simplicity we
try to have as many zero matrix elements as possible. We arrive at a simple
representation, in the basis $\ket{e^{(2)}}
\bra{e^{(1)}},\ket{e^{(2)}}\bra{g^{(1)}},\ket{g^{(2)}}\bra{e^{(1)}},\ket{g^{(2)}}\bra{g^{(1)}}$:
\begin{eqnarray}
E_1&=&\left(\begin{matrix} \frac{C}{\sqrt{1-P_e-B}}& 0\\ 0&\sqrt{1-P_e-B} \end{matrix} \right) \nonumber\\
E_2&=&\left(\begin{matrix} \sqrt{P_e+A-\frac{|C|^2}{1-P_e-B}}& 0\\ 0&0 \end{matrix} \right) \nonumber\\
E_3&=&\left(\begin{matrix} 0& \frac{D^{\ast}}{\sqrt{1-P_e-A}}\\ \sqrt{1-P_e-A}&0 \end{matrix} \right) \nonumber\\
E_4&=&\left(\begin{matrix} 0& \sqrt{P_e+B-\frac{|D|^2}{1-P_e-A}} \\ 0&0 \end{matrix} \right).
\end{eqnarray}
There exists no representation with a smaller number of Kraus operator since we
verified that the rank of the matrix
$(I^{(Q)}\otimes\xi^{(2)})\ket{\beta^{(Q,1)}}\bra{\beta^{(Q,1)}}$,  where
$\ket{\beta^{(Q,1)}}$ is the maximally entangled state
$\ket{\beta^{(Q,1)}}=\frac{1}{\sqrt{2}}(\ket{e^{(Q)},e^{(1)}}+\ket{g^{(Q)},g^{(1)}})$
\cite{Cubi08}, is equal to 4.

\section{Channel capacity}
\label{sec:Holevo}

The classical channel capacity $C$ (often called the product state capacity) of a quantum channel $\xi$ is equal to \cite{Niel00}
\begin{equation}
C(\xi)= \max_{p_j,\rho_j}\left[ S\left(\xi\left(\sum_j p_j \rho_j\right)\right) - \sum_j p_j S\left(\xi\left(\rho_j\right)\right) \right] \label{capacity}
\end{equation}
where S is the Von Neumann entropy $S(\rho):=-Tr(\rho \ln\rho)$.  This quantity corresponds to the amount of reliable classical bit we can send through the quantum channel per use of the channel.

Let us first maximize over the input state to obtain:
$\rho^{(1)}_1=\ket{e^{(1)}}\bra{e^{(1)}}$ and
$\rho^{(1)}_2=\ket{g^{(1)}}\bra{g^{(1)}}$.  The maximization over the
probability $p_1$ gives
\begin{eqnarray}
&&p_1=\frac{2^{w}-P_e-B}{A-B} \nonumber \\
&&w-\ln(1-2^w)=\frac{H(P_e+B)-H(P_e+A)}{A-B}
\end{eqnarray}
where we use the binary entropy $H(p):=-p\ln p - (1-p) \ln (1-p)$.  We finally
arrive at the classical channel capacity $C$, which we divide by $t_f-t_i$ to get
$R$, namely the amount of bits/time which can be sent reliably:
\begin{eqnarray}
R&=& \frac{1}{t_f-t_i}\Bigg[H\left(P_e+p_1 A+\left(1-p_1\right)B\right)\nonumber \\
&& - p_1H\left(P_e+A\right)-\left(1-p_1\right)H\left(P_e+B\right)\Bigg].\label{Cap}
\end{eqnarray}
As expected the classical channel capacity is zero for spacelike interactions
since in that case $A=B=0$.

We remark that the channel capacity as a function of the spacetime separation is a non-analytic function since it identically vanishes outside the lightcone but
is a nontrivial function inside. Any analytic function that vanishes on a
finite interval would of course vanish everywhere. The occurrence of this
non-analyticity may seem surprising since our quantum channel is mapping in
between finite dimensional spaces and therefore appears to be a matter of mere
linear algebra. The non-analyticity arises, of course, from the non-analyticity
of the commutator $[\phi(x),\phi(y)]$ which originates in the fact that, in the
full system, the field lives in an infinite dimensional Hilbert space.
Conversely, if ultraviolet and infrared cutoffs are imposed on the quantum
field theory so that its Hilbert space ${\cal H}^{(3)}$ becomes finite
dimensional, this would reduce these calculations to linear algebra and will
therefore yield some non-vanishing capacity outside the lightcone.
Interestingly, this does not mean that the presence of a natural UV cutoff in
nature would imply a violation of causality. This is because an ultraviolet
cutoff implies that there is in effect a smallest resolvable length, which in
turn means that the very boundaries of the lightcone become unsharp. The
capacity should decay to essentially zero outside the lightcone at a distance
from the lightcone that is about the size of the unsharpness scale induced by
the UV cutoff. Any candidate quantum gravity theory has to reduce to quantum
field theory in a limit and most come with a natural UV cutoff, see e.g.,
\cite{rovelli}. It should be interesting to check causality for such theories
by calculating the channel capacity at distances close to the light cone.

Let us now also consider the quantum channel capacity, \cite{Lloy,Barn}, i.e.,
the amount of quantum information which can reliably be sent through the
channel
\begin{eqnarray}
&&Q(\xi) = \lim_{n\rightarrow\infty} \frac{I_c \left(\xi^{\otimes n}\right)}{n}\nonumber \\
&&I_c(\xi)=\max_{\rho}\left[S\left(\xi\left(\rho\right)\right)-S\left(\xi^{C}\left(\rho\right)\right)\right]
\end{eqnarray}
where $\xi^{C}$ is the complementary channel. For spacelike separated
detections, the quantum channel capacity is zero since the channel is then
anti-degradable: there exists a channel $\Gamma$ such that
$\Gamma\left(\xi^{C}(\rho)\right)=\xi(\rho)$ \cite{Deve}.  This confirms that
superluminal propagation of classical or quantum information is not possible.
For time-like separated detectors, the quantum channel capacity is extremely
hard to compute because the channel is not degradable (a degradable channel is
such that $I_c \left(\xi^{\otimes n}\right)= nI_c \left(\xi\right)$). Indeed, a
theorem in \cite{Cubi08} states that any channel with input and output of
dimension 2 and with Choi rank (minimum number of Kraus operators) bigger than
2 cannot be degradable.  Since the channel we consider has Choi rank equal to
4, it cannot be degradable.  We therefore leave open the question of finding an
explicit expression for the quantum channel capacity of our quantum channel.

\section{Perturbative expansion of the channel}
\label{sec:Pert}

\begin{figure}
    \centering
    \includegraphics[height=0.7cm]{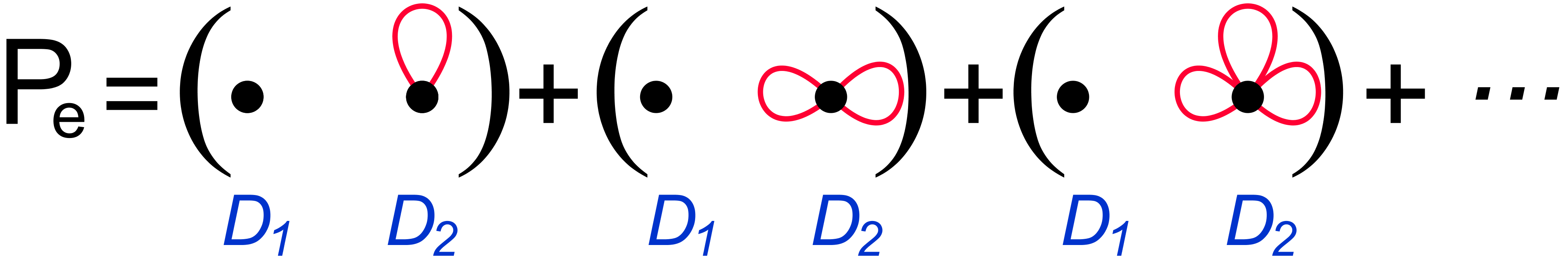}\\
        \includegraphics[height=1.4cm]{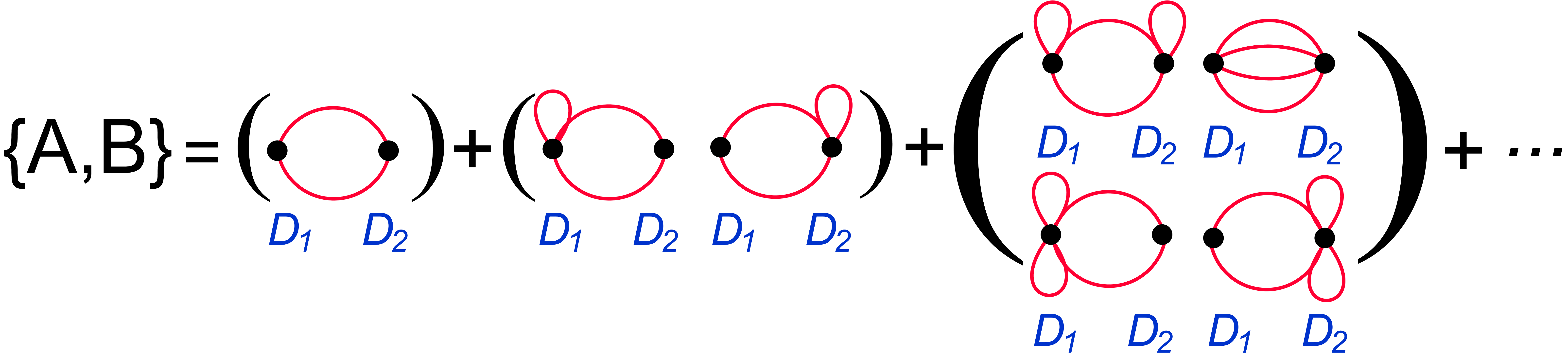}\\
        \includegraphics[height=1.4cm]{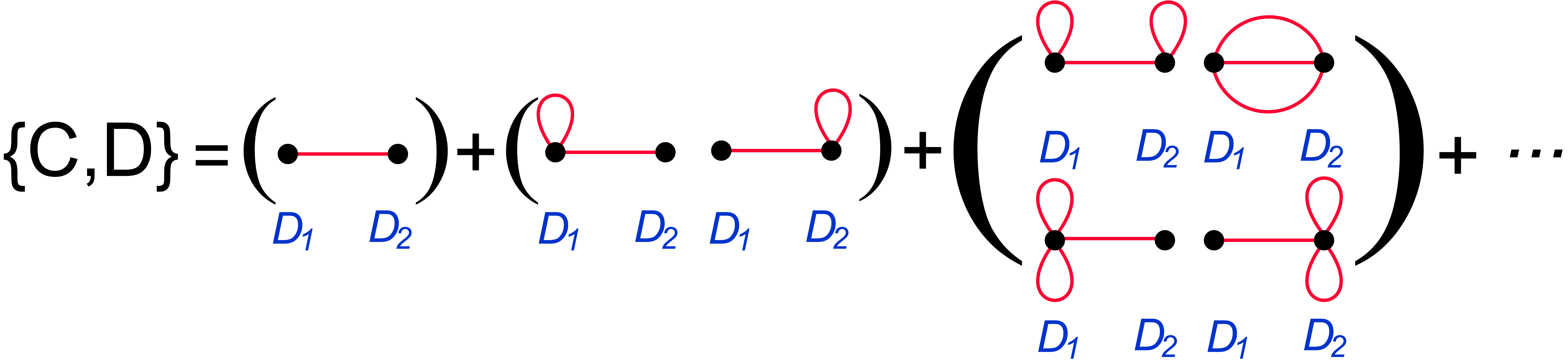}
\renewcommand{\figurename}{FIG.}
    \caption{Feynman diagrams involved in the quantum channel.
$D_j$ stands for detector j, $j\in\{1,2\}$.}
    \label{fig:Feynman}
\end{figure}
Using perturbation theory, we can find explicit expressions for the terms $P_e,
A,B,C$ and $D$ in the weak coupling regime ($\alpha_j\ll 1$).  To this end we use the first orders of the perturbative expansion of equation (\ref{xi}) along
with equation (\ref{channel}), and for simplicity we assume that the field
starts in the vacuum $\ket{0}$:
\begin{widetext}
\begin{eqnarray}
P_e(\Delta E) &=&\alpha_2^2 \int_{t_i}^{t_f} dt_1 \int_{t_i}^{t_f} dt_2 \eta(t_1)\eta(t_2) \bra{0}\phi\left(x_2(t_1)\right) \phi\left(x_2(t_2)\right)\ket{0} e^{-i\Delta E (t_1-t_2)} + O\left(\alpha^4\right)  \label{Pe}\\
A(\Delta E) &=&  2(\alpha_1\alpha_2)^2\int_{t_i}^{t_f}dt_1\int_{t_i}^{t_1}dt_2\int_{t_i}^{t_2}dt_3\int_{t_i}^{t_3}dt_4 \Big\{ \eta(t_1)\eta(t_2)\eta(t_3)\eta(t_4) \cos\left(\Delta E\left(t_1-t_2\right)\right)\nonumber \\
&& \times[\phi\left(x_2(t_1)\right),\phi\left(x_1(t_3)\right)]\Big[ e^{-i\Delta E (t_3-t_4)}\bra{0}\phi\left(x_1(t_4)\right)\phi\left(x_2(t_2)\right)\ket{0} - e^{i\Delta E (t_3-t_4)}\bra{0}\phi\left(x_2(t_2)\right)\phi\left(x_1(t_4)\right)\ket{0}\Big] \nonumber \\
&&+ \left(t_1 \leftrightarrow t_2\right) + \left(t_2 \leftrightarrow t_3\right)+ i\eta(t_1)\eta(t_2)\eta(t_3)\eta(t_4)\sin\left(\Delta E\left(t_2-t_3\right)\right) \nonumber \\
&& \times[\phi\left(x_1(t_2)\right),\phi\left(x_2(t_1)\right)]\Big[ e^{-i\Delta E (t_1-t_4)}\bra{0}\phi\left(x_1(t_3)\right)\phi\left(x_2(t_4)\right)\ket{0}+ e^{i\Delta E (t_1-t_4)}\bra{0}\phi\left(x_2(t_4)\right)\phi\left(x_1(t_3)\right)\ket{0}\Big] \Big\} \nonumber\\
&&+ O\left(\alpha^6\right)   \label{A}\\
B(\Delta E) &=& A(-\Delta E) \nonumber \\
&&+ 4(\alpha_1\alpha_2)^2\int_{t_i}^{t_f}dt_1\int_{t_i}^{t_1}dt_2\int_{t_i}^{t_2}dt_3\int_{t_i}^{t_3}dt_4 \Big\{\eta(t_1)\eta(t_2)\eta(t_3)\eta(t_4) \nonumber\\
&&\times\sin\left(\Delta E(t_2-t_3)\right)\sin\left(\Delta E(t_1-t_4)\right)[\phi\left(x_1(t_2)\right),\phi\left(x_2(t_1)\right)][\phi\left(x_2(t_4)\right),\phi\left(x_1(t_3)\right)] \Big\} + O\left(\alpha^6\right) \label{B}\\
C(\Delta E) &=& \alpha_1\alpha_2 \int_{{t_i}}^{t_f} dt_1 \int_{{t_i}}^{t_1}dt_2 \eta(t_1)\eta(t_2) e^{i\Delta E(t_2-t_1)}[\phi\left(x_1(t_2)\right),\phi\left(x_2(t_1)\right)] + O\left(\alpha^4\right)  \label{C}\\
D(\Delta E) &=& -\alpha_1\alpha_2 \int_{{t_i}}^{t_f} dt_1
\int_{{t_i}}^{t_{1}}dt_2 \eta(t_1)\eta(t_2) e^{i\Delta
E(t_2+t_1)}[\phi\left(x_1(t_2)\right),\phi\left(x_2(t_1)\right)] +
O\left(\alpha^4\right).  \label{D}
\end{eqnarray}
\end{widetext}
We can picture the perturbative expansion with Feynman diagrams \cite{Pesk}, see Fig.(\ref{fig:Feynman}) (the expressions of Eq.(\ref{Pe})-(\ref{D}) are represented by the first diagram of their respective series).   A connection between the two detectors represents a photon emission/absorption process and a connection between a detector and itself (a loop) represents a quantum field fluctuation.  The terms $\{A,B\}$ have an even number of connections between the detectors while the terms $\{C,D\}$ have an odd number of connections.  The only distinction between $A$ and $B$ is the input state at detector 1: the excited state for $A$ and the ground state for $B$.  Thus, the causal connections of $A$ are resonant while the causal connections of $B$ are not resonant.   A similar argument is also true for $C$ and $D$, the connections of $C$ are resonant while the connections of $D$ are not resonant.

\begin{figure}
\includegraphics[height=5cm]{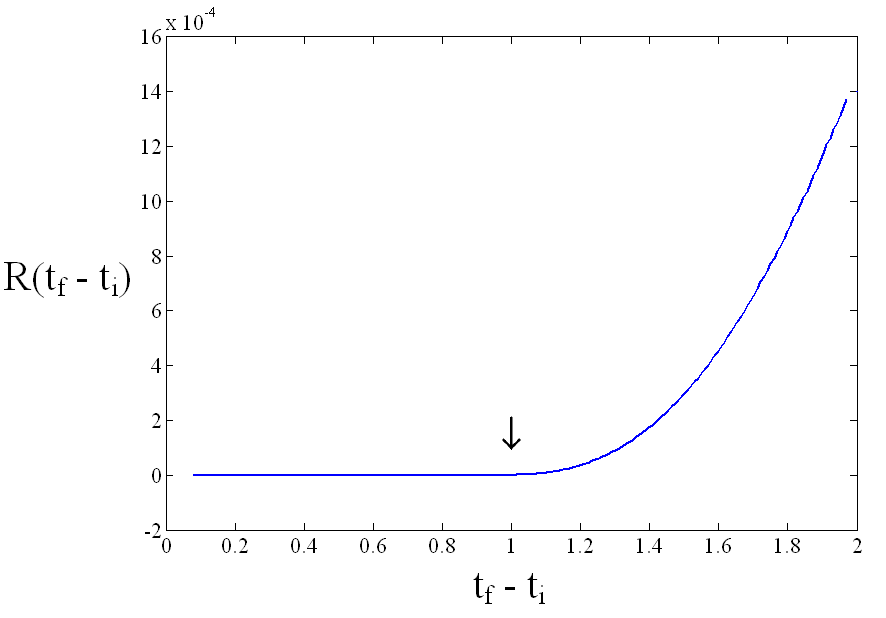}
\renewcommand{\figurename}{FIG.}
\caption{Classical channel capacity as a function of time ($t_f-t_i$) with
$|\vec{x_1}-\vec{x_2}|=1$ and $\Delta E=1$.  The arrow points to the lightcone
$t_f-t_i=|\vec{x_1}-\vec{x_2}|$.} \label{fig:L}
\end{figure}
Using Eq.(\ref{Pe})-(\ref{B}) along with Eq.(\ref{Cap}), we can numerically
evaluate the classical channel capacity as a function of time for inertial
detectors in Minkowsky spacetime, for example, for a massless field, see
Fig.(\ref{fig:L}). The arrow points to the threshold when the spacetime windows
in which the detectors are switched on start to become partially time-like.

\section{Creation of Entanglement in the channel}

Two detectors that interact with a quantum field have access to a renewable
source of entanglement. It has been shown in \cite{Rezn1,Rezn2} that detectors
coupled to a massless quantum field can become entangled even when spacelike
separated. The entanglement was found to appear to propagate in quantum fields
at a speed which depends on the switching functions $\eta(t)$ and on the energy
gap $\Delta E$. The speed of propagation was found to be larger than the speed
of light for suitable $\eta(t)$ and $\Delta E$. A related analysis was also
conducted in an expanding spacetime \cite{Meni}.  In this section we follow up
on these results by showing that the two detectors will in fact automatically
and instantaneously become entangled, namely through what is essentially the
Casimir effect. We find that the Casimir effect entangles significantly which
is encouraging for experimental verification. 

In this section, we switch from the interaction picture to the Schr\"odinger
picture and we assume detectors at rest in Minkowsky spacetime separated by a
fixed distance $\vec{L}:=\vec{x}_2-\vec{x}_1$. This allows us to use perturbation
theory for time-independent perturbations \cite{Coh}. We obtain the new ground
state
\begin{eqnarray}
\ket{e_{g,new}}&=&\ket{e_g}+\sum_{k\neq g}\ket{e_k}\frac{\bra{e_k}H_{int}\ket{e_g}}{E_g-E_k}  \nonumber \\
&& +\sum_{k\neq g}\sum_{l\neq g}\ket{e_k} \frac{\bra{e_k}H_{int}\ket{e_l}\bra{e_l}H_{int}\ket{e_g}}{(E_g-E_k)(E_g-E_l)}\nonumber \\
&& - \frac{\ket{e_g}}{2}\sum_{k\neq g}\frac{|\bra{e_k}H_{int}\ket{e_g}|^2}{(E_k-E_g)^2} + O(\alpha^3) \label{newg}
\end{eqnarray}
where $\ket{e_k}$ are the eigenstates of the  free Hamiltonian and we used the
fact that in our case $\bra{e_g}H_{int}\ket{e_g}=0$.  To regularize the ultraviolet, we give a spatial extent to our detectors:
\begin{eqnarray}
 H_{int}&=&\sum_{j=1}^{2}\alpha_j
\Big(\ket{e^{(j)}}\bra{g^{(j)}}+\ket{g^{(j)}} \bra{e^{(j)}}\Big)\int d^3x f_j(\vec{x})\phi(\vec{x}).\nonumber \\
\end{eqnarray}
Here, the functions $f_j(\vec{x})$ describe the smearing of the detectors, and for simplicity we choose $f_2(\vec{x})=f_1(\vec{x}-\vec{L})$.  Our initial ground state
is $\ket{e_g}=\ket{g^{(1)}}\ket{g^{(2)}}\ket{0}$, and using Eq.(\ref{newg}) the
new ground state $\ket{e_{g,new}}$ is
\begin{eqnarray}
\ket{e_{g,new}}&=&\Bigg[\ket{g^{(1)}}\ket{g^{(2)}}\left(1-\frac{\left(\alpha_1^2+\alpha_2^2\right)}{2}S(\Delta E)\right) \nonumber\\
&&- \alpha_1\ket{e^{(1)}}\ket{g^{(2)}} Q^{(3)}_1(\Delta E)\nonumber\\
&&- \alpha_2\ket{g^{(1)}}\ket{e^{(2)}}Q^{(3)}_2(\Delta E)\nonumber \\
&&+ \alpha_1\alpha_2\ket{e^{(1)}}\ket{e^{(2)}}R(\Delta E,L) +...\Bigg]\ket{0}
\end{eqnarray}
where we use the definitions:
\begin{eqnarray}
Q^{(3)}_j(\Delta E)&:=& \int \frac{d^3p}{(2\pi)^3} \frac{ \int d^3x f_j(\vec{x})e^{-i\vec{p}\cdot\vec{x}}a^{\dag}_{\vec{p}}}{\sqrt{2E_p}(E_p+\Delta E)} \\
R(\Delta E,L)&:=& \int \frac{d^3p}{(2\pi)^3} \frac{e^{i\vec{p}\cdot\vec{L}}\left|\int d^3x f_1(\vec{x})e^{-i\vec{p}\cdot\vec{x}}\right|^2}{2E_p (E_p+\Delta E)(\Delta E)}\nonumber\\&&\\
S(\Delta E) &:=& \int \frac{d^3p}{(2\pi)^3} \frac{\left|\int d^3x f_1(\vec{x})e^{-i\vec{p}\cdot\vec{x}}\right|^2}{2E_p (E_p+\Delta E)^2}.
\end{eqnarray}
The resulting state is clearly entangled because it is a pure state which
cannot be written in a tensor product form.  Let us now ask whether this is
indeed an entangled state from the point of view of the detectors.  To see this
we need to trace out the field, leaving the remaining system in a mixed state
$\rho_{g,new,d}:=Tr_{(3)}\left(\ket{e_{g,new}} \bra{e_{g,new}} \right)$
\begin{eqnarray}
&&\rho_{g,new,d}=\nonumber \\
&&\left(\begin{smallmatrix} 0&0&0&\alpha^2 R(\Delta E, L)\\ 0& \alpha^2S(\Delta E)& \alpha^2 T(\Delta E, L)&0 \\ 0 & \alpha^2 T(\Delta E,L) & \alpha^2 S(\Delta E) & 0 \\ \alpha^2 R(\Delta E,L)&0&0&1-2\alpha^2S(\Delta E)
\end{smallmatrix} \right)\nonumber \\
&&+O(\alpha^4)
\end{eqnarray}
where the matrix is written in the basis $\ket{e^{(1)}e^{(2)}},\ket{e^{(1)}g^{(2)}},\ket{g^{(1)}e^{(2)}},\ket{g^{(1)}g^{(2)}}$, we assumed for simplicity $\alpha_1=\alpha_2=\alpha$
and we use the following definition:
\begin{eqnarray}
T(\Delta E,L) &:=&  \int \frac{d^3p}{(2\pi)^3} \frac{e^{i\vec{p}\cdot\vec{L}}\left|\int d^3x f_1(\vec{x})e^{-i\vec{p}\cdot\vec{x}}\right|^2}{2E_p (E_p+\Delta E)^2}.
\end{eqnarray}

To measure the entanglement of the mixed state, we use the negativity \cite{Vid}, which is twice the absolute value of the sum of the negative eigenvalues of the partial transpose of the density matrix.  We find the negativity $N$ for the density matrix $\rho_{g,new,d}$:
\begin{eqnarray}
N(\Delta E, L)=2\alpha^2\max\left(|R(\Delta E,L)|-S(\Delta E),0\right).
\end{eqnarray}
For simplicity, we analyse this expression when the smearing functions are gaussian
\begin{eqnarray}
f_1(\vec{x})=\frac{e^{-\frac{|\vec{x}-\vec{x}_1|^2}{2{\Delta X}^2}}}{\left( 2\pi\right)^{3/2}{\Delta X}^3}
\end{eqnarray}   
so the size of the detectors is about $\Delta X$.  Such smearing functions could be physically implemented by putting the detectors in a quantum harmonic potential.  Even if gaussian smearing functions have a finite probability for the detectors to overlap, we are only looking at the regime where $L\Delta E \rightarrow 0$, $Lm \rightarrow 0$ and $\frac{L}{\Delta X}\rightarrow \infty$, and in this regime the overlap is insignificant.  In fact, in this regime all the smearing functions have the same effect, namely to create an effective momentum cutoff.  Thus our results would not change for detectors which are delocalized within a region of space which has compact support.  If $\Delta E \gg m$ like for the case of a massless field, we arrive at
\begin{eqnarray}
N\approx \frac{\alpha^2}{2\pi^2}\max\left(\frac{\pi}{2 L \Delta E} -\ln\left(\frac{1}{\Delta E\Delta X}\right) ,0\right).
\end{eqnarray}
Similarly if $\Delta E \ll m$ we have
\begin{eqnarray}
N\approx \frac{\alpha^2}{2\pi^2}\max\left(\frac{\pi}{2 L \Delta E} -\ln\left(\frac{1}{m\Delta X}\right),0  \right).
\end{eqnarray}
We therefore see that the ground state of the interacting theory is entangled from the point of view of the detectors if $L<\frac{\pi}{2\Delta E \ln(1/\Delta E\Delta X)}$ when $\Delta E \gg m$ and if $L<\frac{\pi}{2\Delta E \ln(1/m\Delta X)}$ when $\Delta E \ll m$.  

To estimate how long it takes to extract entanglement from the vacuum, we use the adiabatic theorem.  We assume the system starts in the ground state of the free theory, $\ket{e_g}=\ket{g^{(1)}}\ket{g^{(2)}}\ket{0}$.  Then, the interaction Hamiltonian is smoothly turned on using the switching function $\eta(t)$.  For the system to remain in the ground state, we need $\eta(t)$ to increase slowly enough such that the perturbation is adiabatic.  Following the validity condition for adiabatic behaviour \cite{Saran,Schiff}, we need
\begin{eqnarray}
\max_{t}\left|\frac{\bra{e_k} \dot{H}(t)\ket{e_g}}{E_g(t)-E_k(t)} \right| \ll \min_{t} \left|E_g(t)-E_k(t) \right| \label{adia}
\end{eqnarray}
to hold for any energy level $E_k$.  A rigorous use of the adiabatic theorem requires normalized eigenstates, so let us put our system in a large box of volume $V=L_{IR}^3$.  This procedure creates an infrared cut-off and normalizes the eigenstates of the free Hamiltonian.  Hence, in our case, if we retain only the dominant order, the adiabatic condition translates to:
\begin{eqnarray}
\max_{t}\left|\dot{\eta}(t)\right| &\ll& \left[m^2+3\left(\frac{2\pi}{L_{IR}}\right)^2\right]^{1/4}/\alpha
\nonumber\\
&&\times\left(\sqrt{m^2+3\left(\frac{2\pi}{L_{IR}}\right)^2}+\Delta
E\right)^2. \label{eta} 
\end{eqnarray}
Thus, for a massive field, it is always possible to adiabatically turn on the
interaction, and since the ground state of the interacting theory is entangled,
there will be an instantaneous creation of entanglement.  If the field is
massless, there still is instantaneous creation of entanglement, for any finite
size of box to which we confine our system. Therefore, while Alice and Bob
cannot exchange classical or quantum information faster than the speed of
light, their ability to extract entanglement by interacting with the vacuum is
not bounded by any finite speed.

From Eq.\ref{eta} we notice that in order to obtain the full amount of
entanglement from the ground state,  the system needs an interval of time of
the order of $\left(\max_{t}\left|\dot{\eta}(t)\right|\right)^{-1}$.  This entanglement could either be used
in computations or swapped to other quantum systems for distillation.  After
the entanglement is used up, the detector - field interaction may be switched
off and the system can be put back in the ground state of the free theory
$\ket{e_g}=\ket{g^{(1)}}\ket{g^{(2)}}\ket{0}$, e.g., by cooling. Thus, Alice
and Bob can extract entanglement by interacting with the field in a cyclic and
therefore sustainable way. However, we also see that the extraction of a large
amount of entanglement from the vacuum by this method will cost a large amount
of time. Interestingly, the amount of time needed is determined in a similar
way to how the speed of adiabatic quantum computation is determined.  Recall
that the closeness of eigenvalues determines how fast specific states such as
the ground state can be reached via an adiabatic approach or through cooling,
see e.g. \cite{farhi}. The reason why the finite rate of entanglement
extraction does not lead to a finite speed of entanglement ``propagation", is
that there is no threshold: negativity, indicating entanglement between the
detectors, arises immediately as their interaction with the field is switched
on.

We will now show that the underlying reason why Alice and Bob are entangled
when in the ground state of the interacting theory is that this ground state is
a state in which Alice and Bob are attracted to another through the exchange of
virtual photons. This exchange interaction is in effect the scalar field
version of the Casimir Polder force \cite{Casimir}, which is known to be the
relativistic generalization of the van der Waals force between atoms or
molecules.

Let us now derive the Casimir force between Alice and Bob for point-like detectors $f_1(\vec{x})=\delta(\vec{x}-\vec{x}_1)$. To this end, we
calculate the energy of the new ground state with time independent perturbation
theory \cite{Coh}, and renormalize using $\delta\tilde{E_g}(L):=\delta
E_g(L)-\lim_{L\rightarrow\infty}\delta E_g(L)$. The result of the calculation
is:
\begin{widetext}
\begin{eqnarray}
\delta\tilde{E_g}(L,\Delta E)&=&\sum_{n\neq g}\sum_{k\neq g}\sum_{l\neq g}\frac{\bra{e_g}H_{int}\ket{e_n}\bra{e_n}H_{int}\ket{e_k}\bra{e_k}H_{int}\ket{e_l}\bra{e_l}H_{int}\ket{e_g}}{(E_g-E_n)(E_g-E_k)(E_g-E_l)}+O(\alpha^6)\nonumber\\
&=& -2\alpha^4 \Bigg[ \frac{1}{\Delta E} \Big|\int \frac{d^3p}{(2\pi)^3} \frac{e^{-i\vec{p}\cdot(\vec{x}_1-\vec{x}_2)}}{2E_p(E_p+\Delta E)} \Big|^2 \nonumber \\
&& + \int \frac{d^3p_1}{(2\pi)^3}\int \frac{d^3p_2}{(2\pi)^3}\frac{e^{-i(\vec{p}_1-\vec{p}_2)\cdot(\vec{x}_1-\vec{x}_2)}}{4E_{p_1}E_{p_2}(E_{p_1}+E_{p_2})}\left(\frac{1}{E_{p_1}+\Delta E}+\frac{1}{E_{p_2}+\Delta E}\right)^2  \nonumber \\
&& + 2\int \frac{d^3p_1}{(2\pi)^3}\int \frac{d^3p_2}{(2\pi)^3}\frac{e^{-i(\vec{p}_1-\vec{p}_2)\cdot(\vec{x}_1-\vec{x}_2)}}{4E_{p_1}E_{p_2}(E_{p_1}+E_{p_2}+2\Delta E)(E_{p_1}+\Delta E)(E_{p_2}+\Delta E)} \nonumber \\
&& + \int \frac{d^3p_1}{(2\pi)^3}\int \frac{d^3p_2}{(2\pi)^3}\frac{e^{-i(\vec{p}_1-\vec{p}_2)\cdot(\vec{x}_1-\vec{x}_2)}}{4E_{p_1}E_{p_2}(E_{p_1}+E_{p_2}+2\Delta E)}\left(\frac{1}{(E_{p_1}+\Delta E)^2}+\frac{1}{(E_{p_2}+\Delta E)^2}\right)  \Bigg]+O(\alpha^6). \nonumber\\
\end{eqnarray}
\end{widetext}
The ground state energy is lowered because of the interaction, causing Alice
and Bob to attract each other with the Casimir force $F_C=-\frac{\partial
\delta\tilde{E_g}(L)}{\partial L}$.  For a massless field,
$\delta\tilde{E}(L,\Delta E)\sim\frac{-\alpha^4}{L^4{\Delta E}^3 }$ in the limit
$L \Delta E \rightarrow\infty$. For comparison, the electromagnetic
Casimir-Polder energy \cite{Casimir}, scales as $\sim -L^{-7}$ for large
distances.

Note that so far we did not need to specify the detectors' mass since we
assumed their position to be fixed. Considering now the dynamics of Alice and
Bob due to the Casimir force, it is clear that if their mass is small enough,
their acceleration could be strong enough to become non-adiabatic. In this
case, their motion would cause the system to evolve non-adiabatically and
therefore to become excited. The Casimir force would therefore no longer be
simply the derivative of the ground state energy, $-\frac{\partial
\delta\tilde{E_g}(L)}{\partial L}$ because the state of the system would no
longer be the ground state. To stay in the regime where the Casimir force is
the derivative of the Casimir energy the detectors can move toward each other
at a maximum speed $v$ which needs to be small enough such that the
perturbation is adiabatic. Thus, when our system is in a large box, the
validity condition for adiabatic behaviour of Eq.(\ref{adia}) translates to $v
\ll \frac{{\Delta E}^{3/2}}{\alpha}\frac{32\sqrt{2}}{3\sqrt{3}}$.

\section{Related models}
\label{sec:compare}

A quantum channel modelled by an atom interacting with a photon has recently
been analysed in \cite{Chen08}.  The model uses an atom-photon interaction
given by the Jaynes-Cumming interaction Hamiltonian
$H_{JC}=\alpha\left(\ket{g}\bra{e}\otimes a_{k}^{\dag}+\ket{e}\bra{g}\otimes
a_{k} \right)$ where $a_{k}$ and $a_{k}^{\dag}$ are the annihilation and
creation operator for a single mode $k$.  A similar Hamiltonian was also used
in \cite{Milb} to model a quantized cavity mode kicked by a stream of two-level
atoms.  This interaction Hamiltonian has a natural quantum field
generalization, the Glauber scalar detector \cite{Glau}, which can be used to
model two detectors interacting with a quantum scalar field
\begin{eqnarray}
H_{GS}&=&\sum_{j=1}^{2}\alpha_j\eta\Big(\ket{g^{(j)}}\bra{e^{(j)}}\phi^{-}(x_j)+\ket{e^{(j)}}\bra{g^{(j)}}\phi^{+}(x_j) \Big) \nonumber \label{glauber} \\
\end{eqnarray}
where $\phi^{+}(x)=\int \frac{d^3p}{(2\pi)^3\sqrt{2E_{p}}}e^{-ipx}a_{\vec{p}}$
and $\phi^{-}(x)=\phi^{+\dag}(x)$ are respectively the positive and negative
frequency part of the field.  While this detector is not sensitive to the
quantum fluctuations of the field, i.e., in our notation, $P_e=0$, this
detector model allows non-local effects, see \cite{Busc09}. We can confirm the
non-locality by using in our channel Glauber detectors instead of Unruh-DeWitt
detectors. To this end, we use Eq.\ref{glauber} in Eq.(\ref{xi}). We see that
then terms that are dependent on $\rho^{(1)}$ are no longer necessarily
proportional to $[\phi(x_1),\phi(x_2)]$. Using the perturbative expansion of
the channel in Sec.V shows that non-causal terms appear already in the
$O(\alpha^2)$ order:
\begin{eqnarray}
\xi(\rho^{(1)})&=&\ket{g^{(2)}}\bra{g^{(2)}}\nonumber\\
&&-\alpha_1\alpha_2\int_{t_i}^{t_f}dt_1\int_{t_i}^{t_1}dt_2\Big[\eta(t_1)\eta(t_2)\nonumber\\
&&\times e^{i\Delta E(t_1-t_2)}  \ket{e^{(2)}}\bra{g^{(2)}}\bra{e^{(1)}}\rho^{(1)}\ket{g^{(1)}}\nonumber\\
&&\times D\left(x_2\left(t_1\right)-x_1\left(t_2\right)\right)+c.c.\Big]+O\left(\alpha^4\right). \nonumber \\
\end{eqnarray}
Here, $D\left(x-y\right):= \bra{0} \phi^{+}(x)\phi^{-}(y)\ket{0}$. Since the
correlator $D\left(x-y\right)$ is not vanishing outside the lightcone, detector
2 would indeed be influenced by detector 1 as soon as the interaction is turned
on even if the detectors are spacelike separated.  It may be interesting to see if similar effectively non-local detectors, such as the one in \cite{Piazza}, behaves causally or non-causally under our channel picture.  

\section{Outlook}

The type of quantum channel that we here considered could be useful, for
example, in the context of implementations of quantum networks, where photons
carry quantum information in between atoms that possess effectively two levels,
see e.g., \cite{Cira97}. But it should also be straightforward to generalize
our study to detectors with any number of energy levels. The number and spacing
of the energy levels of the detectors should translate into an effective
alphabet size. This should also allow one to generalize the results of
\cite{bowen}, where it was first shown how quantum noise imposes a natural
bound to the capacity of an otherwise noiseless bosonic channel. The analysis
of \cite{bowen} employed the time-energy uncertainty principle to describe the
limit to the distinguishability of photons of energy difference $\Delta E$ in
an observation time $\Delta t$. It should be interesting to re-analyze these
results within the present information-theoretic framework of the quantum
channel in which all effects of quantum noise are built in from the start.

It should also be interesting to generalize our model to yield a new approach
to analyzing the setup of \cite{Hayd}, where Alice and Bob are inertial
observers which are exchanging modes of a quantum field, while Eve is
accelerating and tries to intercept the message.  It was shown there that,
because of the Unruh effect, it is always possible for Alice and Bob to
communicate privately.  To show this, the approach to the Unruh effect using
Bogoliubov transformations was used. Generalizing our setup, one may use
Unruh-DeWitt detectors, which are known to allow a more flexible description of
the Unruh effect. For example, Eve would not have to accelerate uniformly and
could indeed take an arbitrary trajectory.

The channel which we studied here should also be generalizable to curved
spacetimes to study, for example, the impact of spacetime expansion and
horizons. Finally, let us recall that, in the presence of a suitable natural
ultraviolet cutoff, the density of degrees of freedom in quantum fields is
finite, see e.g. \cite{Kempf}. It should be interesting to investigate how this
finite density of degrees of freedom translates into a finite information
carrying capacity of quantum fields, in the concrete sense of the capacity of
quantum channels. Indeed, the quantum channel that we investigated here can be
interpreted as describing one detector which imprints information in a quantum
field, and a second detector reading out this information. The approach
therefore allows one to ask questions such as, how write and read cycles can be
optimized, how much information is left in the field after a cycle, or how much
quantum or classical information can maximally be written into and retrieved
from a quantum field in some finite region of spacetime.
\\

\section*{Acknowledgments}

The authors whish to thank Ralf Sch\"utzhold for valuable comments and suggestions.  M.C. acknowledges support from the NSERC PGS program. A.K. acknowledges support from CFI, OIT, the Discovery and Canada Research Chair programs of NSERC and is grateful for the very kind hospitality at the University of Queensland during the early stages of this work.

\end{document}